\documentclass[prl,a4paper,twocolumn,superscriptaddress,showpacs,floatfix]{revtex4}
\usepackage[dvips]{graphicx}
\usepackage{amsfonts}
\usepackage{amssymb}

\begin{document}

\title{Disorder induced collapse of the electron phonon coupling in MgB$_{2}$ observed by Raman Spectroscopy}
\date{\today}
\author{K.A. Yates}
\email[]{kay21@cam.ac.uk}
\author{G. Burnell}
\author{N.A. Stelmashenko}
\author{D.-J. Kang}
\affiliation{Department of Materials Science, University of Cambridge, Cambridge, CB2 3QZ, UK}
\author{H.N. Lee}
\author{B.Oh}
\affiliation{L.G. Electronics Institute of Technology, Seoul 137-724, Korea}
\author{M.G. Blamire}
\affiliation{Department of Materials Science, University of Cambridge, Cambridge, CB2 3QZ, UK}

\begin{abstract}
The Raman spectrum of the superconductor MgB$_{2}$ has been measured as a function of the Tc of the film.  A striking correlation is observed between the $T_{c}$ onset and the frequency of the $E_{2g}$ mode.  Analysis of the data with the McMillan formula provides clear experimental evidence for the collapse of the electron phonon coupling at the temperature predicted for the convergence of two superconducting gaps into one observable gap.  This gives indirect evidence of the convergence of the two gaps and direct evidence of a transition to an isotropic state at 19 K.  The value of the electron phonon coupling constant is found to be 1.22 for films with T$_{c}$ 39K and 0.80 for films with T$_{c}\leq$19K.
\end{abstract}
\maketitle

Since the discovery of superconductivity in MgB$_{2}$ in January 2001 \cite{Nagamatsu} a large volume of work has been directed towards developing and improving the quality of MgB$_{2}$ thin films \cite{Uedarev}. The temperature of the onset of superconductivity ($T_{c}$) in these films has varied considerably, although there is no consensus on the mechanism behind the $T_{c}$ suppression.  Nonetheless, theoretical models suggest that a two band superconductivity model is needed to explain the high $T_{c}$ in MgB$_{2}$ \cite{Kortus, Mazin, Liu, Brinkman}.  As inter-band scattering increases due to impurities in the boron layer, so the two gaps in MgB$_{2}$ are predicted to converge into one gap with $\Delta _{\sigma} = \Delta _{\pi}$ = 4.1 meV \cite{Brinkman}; this transition is predicted to occur as the $T_{c}$ is suppressed below 25 K \cite{Brinkman}.  Measurements of the phonon energies in YBa$_{2}$Cu$_{3}$O$_{7-\delta}$ helped the understanding of the relation between the crystal structure and the $T_{c}$ in this material \cite{Gibson}.  Therefore, similar measurements made on MgB$_{2}$ films with different $T_{c}$s may help to elucidate the mechanisms behind the suppression of the critical temperatures in MgB$_{2}$,  while casting light on the nature of the two band superconductivity.  

Numerical calculations indicate that in MgB$_{2}$  there are two phonon modes involving vibrations along the c-axis   ($A_{2u}$ and $B_{1g}$) and two doubly degenerate modes with vibrations only in the ab plane ($E_{1u}$ and $E_{2g}$) \cite{Yildirim}; for the \textit{P$_{6}$mmm} space group to which MgB$_{2}$ belongs, only the $E_{2g}$ mode is Raman active \cite{Kunc}.  Predictions for the frequency of the $E_{2g}$ mode vary between 515 cm$^{-1}$ (64 meV) \cite{Kortus} and 665 cm$^{-1}$ (82 meV) \cite{Satta}, while experimental studies have shown a broad Raman mode between 600 cm$^{-1}$ and 630 cm$^{-1}$ \cite{Kunc,Postorino,Goncharov,Bohnen,Chen,Hlinka,Meletov,Quilty2,Rafailov}. The consensus is that this mode is the $E_{2g}$ mode, however, there remains controversy about this assignment \cite{Kunc,Chen}, with Kunc \textit{et al.} \cite{Kunc} concluding that this mode was due to a ``contaminant phase at the sample surface''. In this letter Raman spectroscopy has been used to characterise the quality of MgB$_{2}$ films both in terms of impurity phases and disorder. The effects of a suppressed $T_{c}$ on the phonon mode at $\sim$600 cm$^{-1}$ were studied. Both magnesium stoichiometry and oxygen contamination may play a role in determining the $T_{c}$ \cite{Eom,Serquis}. It has been shown previously that as-grown films of MgB$_{2}$ can have significant magnesium concentration gradients through the thickness of the film \cite{Jo}.  This effect is equally likely where a precursor boron film has been annealed in magnesium vapor. Therefore, measurement of the superconducting properties of a post-annealed boron film through the thickness of the film would give an indication as to the effect of magnesium stoichiometry on the $T_{c}$.

The films used in this study were prepared by two different techniques to produce films with different $T_{c}$s. The samples  were either grown by DC magnetron sputtering from separate Mg and B targets onto c-sapphire substrates at room temperature, followed by an anneal at varying temperatures and times with the substrates face down on the heater \cite{Nadia}, or they were produced by ex-situ annealing of evaporated boron films in magnesium vapor \cite{Koreans}. X-ray analysis showed that the ex-situ annealed films had c-axis orientation \cite{Koreans} however, no evidence of any orientation was found in the sputtered films.  Atomic force microscopy (AFM) indicated that over the sputtered films there was little change in crystallite size with the average crystallite measuring $\sim$ 100 nm.  In comparison, the ex-situ films had an average crystal size less than 100 nm. An ex-situ annealed boron film of nominal thickness 500 nm and $T_{c}$ of $\sim$39 K was then formed into a staircase structure with step heights of $\sim$30 nm by a standard lithography technique followed by broad-beam Ar ion milling (500 V, 1 mA cm$^{-2}$ beam), shown schematically in the inset to Fig. \ref{Armill}. Raman spectra taken before and after the first milling process did not show any appreciable effect due to the milling process or the aqueous photoresist developer used and so the effects of the patterning process were taken to be negligible.

\begin{figure}
\begin{center}
\includegraphics[width=7.5cm]{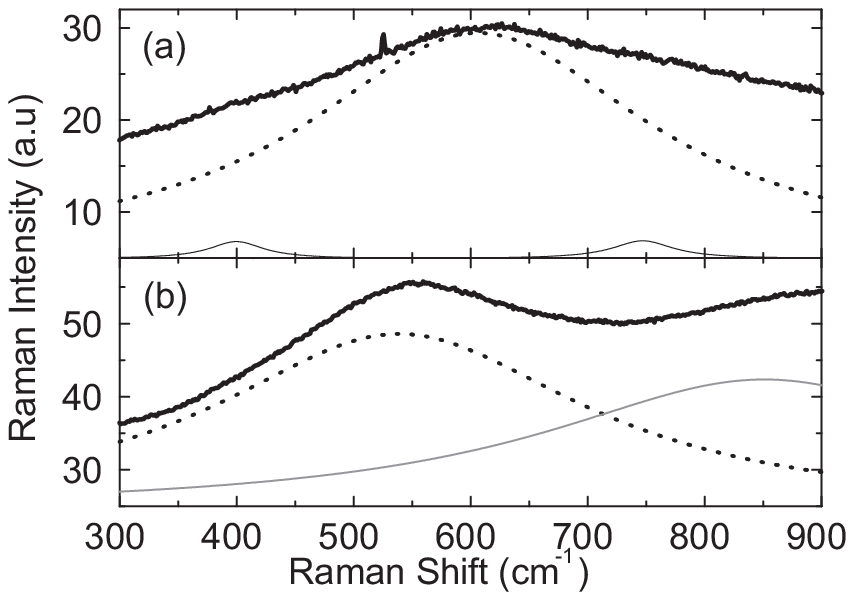}
\caption{\label{spec}}
\end{center}
\end{figure}

Raman spectra were taken using a Renishaw Ramascope 1000 with an Ar ion laser at 514nm at room temperature.  The laser spot size was 3$\mu$m x 3$\mu$m. In order for the background removal to be systemmatic across the series, the background removal method of Postorino et al, was used \cite{Postorino} which is fully described in \cite{Postorino} and references therein. Error bars on the phonon frequency were estimated from fitting several spectra from different parts of the same film and calculating the standard deviation.  Resistance data were taken by a normal four point contact technique.  The Raman spectra of films with $T_{c}\sim$38 K could be fitted well by a broad (FWHM $\simeq$180 cm$^{-1}$) Lorentzian curve centred at 611 cm$^{-1}$, with some evidence of secondary peaks at 400 cm$^{-1}$ and $\sim$ 700 cm$^{-1}$, previously associated with disorder in the structure of the MgB$_{2}$, \cite{Postorino,Rafailov}, see Fig. \ref{spec}.  Figure \ref{spec}b shows the Raman spectrum of a film with a $T_{c}$ of 12.5K, where the frequency of the main mode has decreased to $\simeq$535 cm$^{-1}$.  There is also evidence of additional, higher frequency contributions, the origins of which are unknown but may include contributions from B$_{2}$O$_{3}$ and will be discussed in a future publication.

\begin{figure}
\begin{center}
\includegraphics[width=7.5cm]{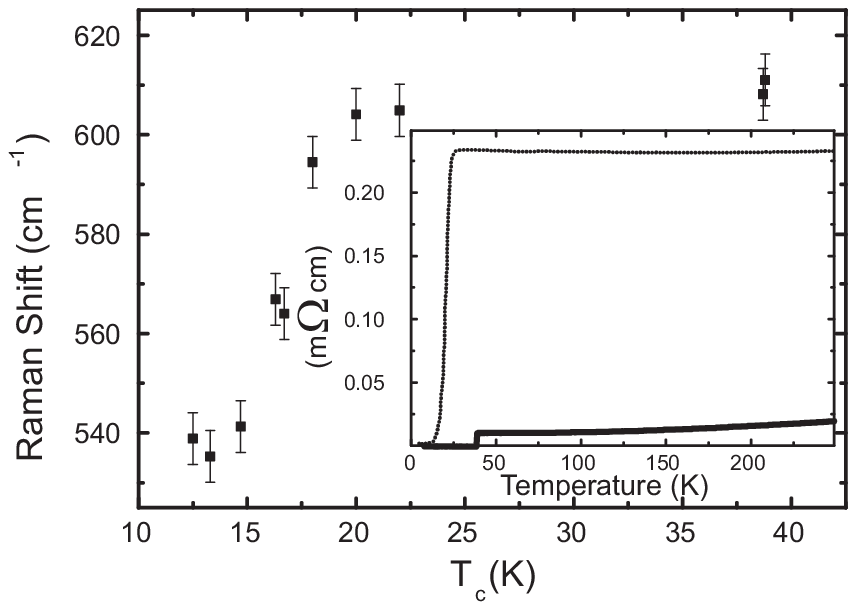}
\caption{\label{Tcplot}}
\end{center}
\end{figure}

\begin{figure}
\begin{center}
\includegraphics[width=7.5cm]{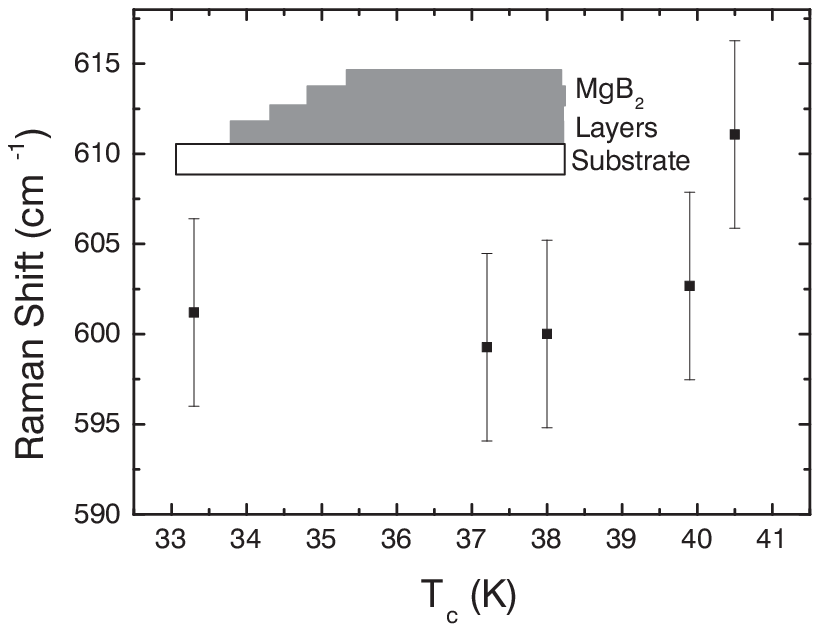}
\caption{\label{Armill}}
\end{center}
\end{figure}

The main panel of Fig. \ref{Tcplot} shows the frequency of the $E_{2g}$ mode as a function of $T_{c}$. As the $T_{c}$ of the films decreases from 38 K to 22 K, the frequency of the $E_{2g}$ mode decreases  from 615 cm$^{-1}$ to $\sim$600 cm$^{-1}$, values similar to the range reported in \cite{Kunc, Postorino, Goncharov, Bohnen}.  As the $T_{c}$ of the films is suppressed below 20 K, there is a sudden large drop in the frequency of this mode from 600 cm$^{-1}$ to 540 cm$^{-1}$ (7 meV) over an 8 K drop in $T_{c}$. The inset to Fig. \ref{Tcplot} shows the resistivity as a function of temperature in two films of very different onset $T_{c}$.

Figure \ref{Armill} shows the frequency of the $E_{2g}$ mode through the thickness of the film L1.  The frequency of this mode drops in the first 150 nm of the surface of the film from 611 cm$^{-1}$  to $\sim$600 cm$^{-1}$ whereupon it stays approximately constant throughout the film thickness.  
The dependence of the $E_{2g}$ mode on the $T_{c}$ indicated in Fig. \ref{Tcplot} is again seen in the staircase film. RT data was taken on the staircase film by means of wire bonding contacts to each individual step and measuring the voltage drop across each step individually, in sequence, as a function of temperature. Resistance vs temperature measurements indicated a region of activated transport immediately above the superconducting transition temperature across the second and third thinnest steps, reminiscent of some of the films with depressed $T_{c}$s both in Fig. \ref{Tcplot} and in the literature, see for example \cite{Eom}.

\begin{figure}
\begin{center}
\includegraphics[width=7.5cm]{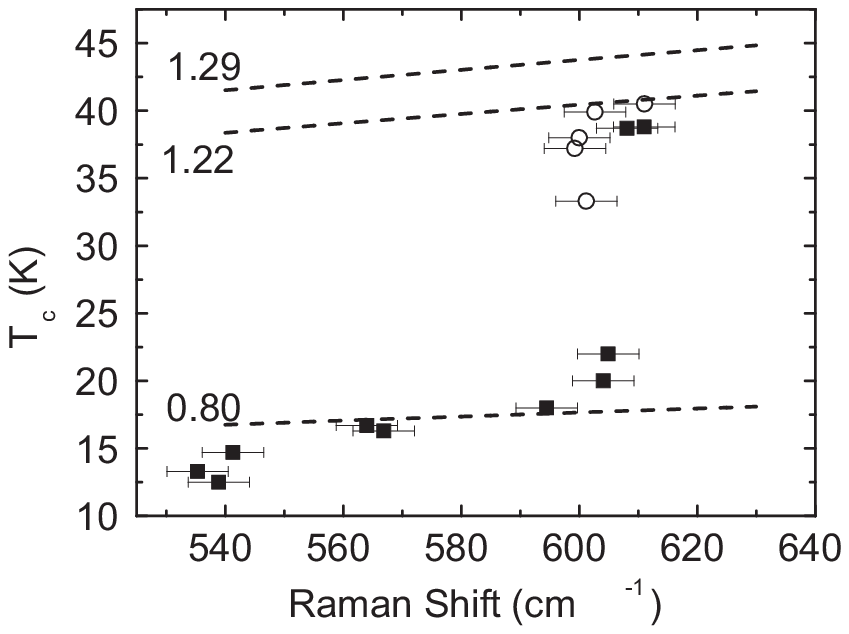}
\caption{\label{ther}}
\end{center}
\end{figure}

The strong dependence of the frequency of the mode detected at around 600 cm$^{-1}$ on the $T_{c}$ of the film suggests that an internal mechanism is at work for the suppression of $T_{c}$ in these films and would appear to rule out this mode being a surface contaminant.  In order to understand the relationship between the phonon frequency and the $T_{c}$, the McMillan formula \cite{McMillan} may be used:
\begin{equation}
T_{c}=\frac{\left<\omega _{log}\right>}{1.2}\exp \left[-1.02\frac{1+\lambda}{\lambda - \mu ^{*}(1+\lambda)}\right]
\end{equation}
where, following the method of ref. \cite{Kortus} we have taken $\left<\omega _{log}\right>=(717\times \omega ^{2}_{E_{2g}} \times 435)^{0.25}$ with 717 and 435 being the phonon frequencies of other modes in the MgB$_{2}$ system taken from ref. \cite{Osborn}. $\mu ^{*}$ is the Coulomb pseudopotential, taken as equal to 0.13 \cite{Brinkman}, and $\lambda$ is the electron phonon coupling constant.  For single crystal ($T_{c}$ = 38K) MgB$_{2}$, $\lambda _{\sigma \sigma}$=1.29 \cite{Cooper}.  Taking these values and $\omega _{E_{2g}}$ = 620 cm$^{-1}$ gives $T_{c}$ = 43 K.  For the films of lower $T_{c}$ a lower value of $\lambda$ must be used if the $T_{c}$ found from Eq. 1 is to agree with that found experimentally.  In films of high Tc with little interband scattering, the effective coupling parameter is given by the maximum eigenvalue of the matrix given by the interaction parameters multiplied by the density of states in that band \cite{Liu}.  For MgB$_{2}$ this is closer to $\lambda _{\sigma \sigma}$ than to $\lambda _{\pi \pi}$ \cite{Liu}.  However, in the limit of strong interband scattering, so the effective coupling parameter becomes a weighted average over all of the bands \cite{Golubov2} and is therefore reduced.  Although the use of the McMillan formula using an average value of $\lambda$ has been questioned \cite{Choi}, it has been used successfully in other studies where a higher value of $\lambda$ is used \cite{Kortus, Loa}.  A simple average gives $\lambda$ = 0.458, which leads to a $T_{c}$ ($\omega _{E_{2g}}$ = 540 cm$^{-1}$) of 1.6 K.  If, instead, a value of $\lambda$ = 0.80 is taken a much better fit is obtained to the data, see Fig. \ref{ther}.  The data may be fitted very well assuming $\lambda$ is 1.22 for the high $T_{c}$ films and  0.80 for the films with lower $T_{c}$.  Although the model used is simple and hence accurate determination of the electron phonon coupling cannot be assumed, the results roughly correspond with those obtained via the de Haas van Alphen effect \cite{Cooper}.  Particularly important though is that the electron phonon coupling must drop substantially at $\simeq$ 19 K, indicating that a fundamental change to the nature of the superconducting state is occuring.
Such a shift in the value of the electron-phonon coupling would be expected if the superconducting state became isotropic and the two gaps merged into one superconducting gap. The point at which this occurs may be estimated from the point at which the data can be fitted by a lower range of $\lambda$.  Comparing Fig. \ref{ther} with Fig. \ref{Tcplot} indicates that this occurs at $T_{c}$ = 19$\pm$1 K, in satisfactory agreement with the predicted value of 25 K \cite{Brinkman} and excellent agreement with the 19 K value given by Choi et al. \cite{Choi} and Mazin et al, \cite{MazinPhysC}. 

The $T_{c}$ data from the thick film made into a staircase structure indicate that the physical mechanism behind the $T_{c}$ suppression may be in part attributable to magnesium stoichiometry.  An additional factor that may arise particularly in the lower $T_{c}$ films is that of oxygen alloying into the MgB$_{2}$ film \cite{Eom}. As well as changing the lattice parameters and, therefore, the phonon frequencies, an additional effect of oxygen alloying into the boron plane would be to introduce defects into the boron layer thereby causing the increase in interband scattering which leads to the collapse of the two gaps into one \cite{Kortus,Liu,Mazin,Brinkman}.  It is therefore possible that the reduction of $T_{c}$ in this region is caused by such a mechanism with the frequency of the $E_{2g}$ mode reflecting the presence of the oxide impurities within the boron layers.  Although the Raman spectrum of an Mg$_ {x}$B$_{y}$O$_{z}$ compound is unknown, a symptom of such alloying may be the presence of B$_{2}$O$_{3}$ in the grain boundary regions \cite{Klie}.  Taking the ratio of the intensity of the 808 cm$^{-1}$ B$_{2}$O$_{3}$ mode \cite{RamanB2O3} to the $\simeq$600 cm$^{-1}$ mode of MgB$_{2}$ gives an indication as to the relative abundance of the two compounds.  As the $T_{c}$ drops, the ratio of oxide to MgB$_{2}$ content increases, however, such analysis needs very careful consideration as no work has yet been done on the intensity of the 600 cm$^{-1}$ mode as a function of structure and it is not unfeasible that it is varying independently of the amount of oxide in the film. 

In conclusion, we have shown that the mode at $\sim$600 cm$^{-1}$ in MgB$_{2}$ is strongly correlated with the $T_{c}$ which would support the hypothesis that this mode is the sought after $E_{2g}$ mode in MgB$_{2}$.  Raman spectra taken as a function of thickness through a thin film allowed a prediction of a $T_{c}$ drop between the surface layer and the substrate layer which was later observed by transport data taken across the film.  The reduction in $T_{c}$ observed in MgB$_{2}$ was attributed to a mixture of magnesium off-stoichiometry and oxygen alloying into the boron layers. However, on its own, the drop in frequency of the phonon mode is not enough to explain the drop in $T_{c}$ that is observed.  Such a decrease may only be explained if the electron phonon coupling constant changed abruptly between 19 and 25 K from $\lambda$ $\simeq$ 1.22 for films with $T_{c} >$ 19K to $\lambda$ $\simeq$ 0.8 for $T_{c} \leq$ 19K, consistent with the expected transition from a two gap superconductivity in MgB$_{2}$ samples with high values of $T_{c}$ to one in which there is only one, isotropic, superconducting gap for samples that have high interband scattering and a reduced value of $T_{c}$.

\begin{acknowledgments}
This work was supported by the EPSRC.  The authors are grateful for useful discussions with Y. Bugoslavsky, L.F. Cohen and J.R. Cooper.
\end{acknowledgments}

\bibliography{Raman}

\newpage

\large{Figure Captions}\\

\normalsize{
FIG. 1. Raman spectra (thick solid line) of (a) a film with $T_{c}$ = 38K and (b) with $T_{c}$ = 12.5K, the $E_{2g}$ mode fit is shown as the dotted line in the figure, also shown are additional contributions at 400 cm$^{-1}$ and 700 cm$^{-1}$ in (a), while the grey line in (b) is a fit to a peak of unknown origin.

FIG. 2. Wave number of the $E_{2g}$ mode for films of different $T_{c}$, inset shows the resistivity as a function of temperature for two films across the series.

FIG. 3. Wave number of the $E_{2g}$ mode as a function of the $T_{c}$ through the thickness of film L1, inset shows the staircase structure of the film.

FIG. 4. The $T_{c}$ as a function of phonon frequency for the thin films with different $T_{c}$ \footnotesize{($\blacksquare$)},\normalsize{ the staircase structure film }(\small{$\circ$)}\normalsize{ and the fit to the data using the McMillan formula with $\lambda$ = 0.8,1.22,1.29 (dashed lines)}.
}
\newpage

\end{document}